\documentstyle[preprint,aps]{revtex}
\begin{document}
\draft
\preprint{\vbox{
\hbox{CTP-TAMU-41/94}
\hbox{TRI-PP-94-71}
\hbox{hep-ph/9408375}
\hbox{August 1994}
}}

\title{
The effect of new physics on $R_b=\Gamma(Z\to b\overline{b})/
\Gamma(Z\to\hbox{hadrons})$
}

\author{James T. Liu}
\address{
Center for Theoretical Physics, Department of Physics\\
Texas A\&M University, College Station, TX 77843--4242
}
\author{Daniel Ng}
\address{
TRIUMF, 4004 Wesbrook Mall\\
Vancouver, B.C., V6T 2A3, Canada
}
\maketitle
\begin{abstract}
Motivated by the $\sim 2\sigma$ discrepancy between the experimental value
for $R_b$ and the theoretical prediction of the Standard Model, we examine
the effect of new physics on the $Z$-$b$-$\overline{b}$ vertex.  Using
general results for both new scalars and gauge bosons, we show that two
conditions must be satisfied in order for new physics to give observable
deviations in the vertex.  In particular, the fermion in the loop must
transform chirally under $SU(2)\times U(1)$ and it must be massive compared
to the exchanged boson.  We examine the implications of these results
on the 2 Higgs Doublet model and the Left-Right Symmetric model.
\end{abstract}

\narrowtext

\section{Introduction}
In the last few years, precision electroweak experiments have improved
to the point where they are now providing highly sensitive tests of the
Standard Model (SM).  While the SM is generally in excellent agreement with
experiment, recent results on the left-right asymmetry $A_{LR}$ at SLC
\cite{slcalr} and $R_b=\Gamma(Z\to b\overline{b})/\Gamma(Z\to
\hbox{hadrons})$ measured at LEP \cite{rbref} indicate a possible
disagreement at the 2 to 2.5 $\sigma$ level.  Although the discrepancy
between $A_{LR}$ and the LEP asymmetry results is difficult to accommodate,
$R_b$ presents the interesting possibility that new physics beyond the SM
may show up non-obliquely.

A common approach to studying new physics is to assume that the dominant
effect comes from oblique corrections.  In this manner, deviations from the
SM may parametrized, for example, in terms of the parameters $S$, $T$ and
$U$ \cite{peskin} or $\epsilon_{N1}$, $\epsilon_{N2}$ and $\epsilon_{N3}$
\cite{altarelli}.  Contributions from box and vertex diagrams with heavy new
particles are typically small.  However, in the context of the SM, the
$Z$-$b$-$\overline{b}$ vertex receives an important contribution from heavy
top loops \cite{akhundov,bernabeu,beenakker}, leading us to speculate
whether new physics may also play a role in direct corrections.

While all fermion vertices may receive corrections from new particles, we
choose to focus on the $b$ vertex.  After all, in the SM, this is the
only vertex that receives a large correction.  We take the heavy top as a
hint that the third family is somehow singled out, and that this
characteristic may persist even beyond the SM.  For instance, in the 2 Higgs
Doublet (2HD) model, while the additional physical Higgs particles will lead
to new vertex contributions, the only ones of importance lie in the
third family, as it is the only family with significant Yukawa couplings.
Nevertheless, our analysis is easily generalized to include vertex
corrections to the first two families as well.

The ratio $R_b$ provides an excellent test of the $b$ vertex both because
it is fairly insensitive to QCD corrections and because the oblique
corrections are mostly cancelled in the ratio.  The latter allows us to
focus solely on the $b$ vertex without worrying about the effect of new
physics on the oblique parameters.  Experimentally, the LEP collaborations
have measured $R_b$ to be $R_b=0.2208\pm0.0024$ \cite{rbref}, which disagrees
with the theoretical value of $\approx 0.215$ for $m_t\approx174$~GeV.  In
this letter, we explore whether vertex corrections from new physics can
bring the theoretical predictions closer to the experimental result.  In
particular, we examine the effect of both new scalars and new gauge bosons
on the $b$ vertex and hence their impact on $R_b$.  We find that, although
large contributions are possible from new physics, they depend on the
presence of massive chiral fermions in the vertex --- either the top or
possible new fermions.

% vector ward identity when q^2->0

\section{The ratio $R_{\lowercase{b}}$ and the ${\lowercase{b}}$ vertex}
We work in the * scheme \cite{kennedy}, where the partial width of the
$Z$ into fermion pairs is given by
\begin{equation}
\Gamma(Z\to f\overline{f})={\alpha_*\over6s_*^2c_*^2}N_c\beta M_Z Z_{Z*}
\left[(1-x)((a_L^f)^2+(a_R^f)^2)+6xa_L^fa_R^f\right]\ ,
\end{equation}
where $x=(m_b/M_Z)^2$ and $\beta=\sqrt{1-4x}$.  $N_c$ is the color factor
and $a_L^f$ and $a_R^f$ are the left- and right-handed fermion couplings
to the $Z$ gauge boson.
Since we take the ratio of widths to get $R_b$, most oblique and QCD
corrections cancel.  Thus we need not worry about effects such as the
propagator renormalization hidden in $Z_{Z*}$.  The weak mixing angle
in the * scheme, $s_*$, will pick up oblique corrections.  However such
effects turn out to be fairly small.

At tree-level, the couplings are given simply by $a_L^f=T_3(f)-Q(f)s_*^2$ and
$a_R^f=-Q(f)s_*^2$.  However, even in the context of the minimal SM, the
experimental data is accurate enough that it is important to take SM loops into
account.  Since we are interested in additional vertex corrections beyond
the SM, we separate out the new physics and parametrize the left- and
right-handed couplings by
\begin{equation}
a_{L,R}^f=a_{L,R}^{f,\rm SM}+{\alpha_*\over4\pi s_*^2}
\delta a_{L,R}^f\ ,
\end{equation}
where $a_{L,R}^{f,\rm SM}$ include the SM vertex corrections%
\footnote{To be general, the dipole form factors should also be included.
However, in the absence of left-right mixing, they are suppressed by a
factor of $m_b/M_W$ and may be ignored.  Even when mixing is present, the
dipole contributions are often suppressed, as we will point out later.}.
For $f\ne b$,
the corrections are small but non-negligible.  However, for
the $b$ vertex, $a_L^b$ receives an important non-universal contribution due
to heavy top loops.  Although there is now evidence for a top of mass
$m_t=174\pm 17$ Gev \cite{cdftop}, we wish to leave the top mass as a free
parameter so we may study the effects of $m_t$ on the vertex.  Hence we
take $a_L^{b,\rm SM}\equiv a_L^{b,\rm SM}(m_t=0)$ and instead
incorporate $m_t$ into $\delta a_L^b$ according to
\begin{equation}
\delta a_L^b = [a_L^{b,\rm SM}(m_t)-a_L^{b,\rm SM}(m_t=0)]+\cdots
\label{eq:alsm}
\end{equation}
where $\cdots$ signifies the contribution from new physics.

To be precise, we note that in general the SM vertex correction may have
imaginary parts due to both $Z$ and $W$ loops.  However, since we know
$m_t>M_Z/2$, this imaginary part arises only from the $Z$ loop and may be
incorporated entirely in $a_L^b$ itself.  We always assume this is done so
that $\delta a_L^b$ is real.  Furthermore, since we expect any new
particles to be heavy, both $\delta a_L^b$ and $\delta a_R^b$ will remain
real in the presence of new physics.

For the SM prediction of $R_b$, we take $R_b^{\rm SM}(m_t=0)=0.2179\pm0.0004$
where the uncertainty mainly arises from the determination of $m_b$.  Then
to linearized order, and ignoring the $b$-quark mass, $\delta a_{L,R}^b$
shift this prediction according to
\begin{eqnarray}
R_b&=&R_b^{\rm SM}+R_b^{\rm SM}(1-R_b^{\rm SM}) {\alpha_*\over2\pi s_*^2}
{a_L^b\delta a_L^b+a_R^b\delta a_R^b\over (a_L^b)^2+(a_R^b)^2}\nonumber\\
&\approx&0.2179-0.0021\,\delta a_L^b+0.00038\,\delta a_R^b\ .
\label{eq:rb}
\end{eqnarray}
In the SM, $\delta a_R^b=0$, whereas $\delta a_L^b\sim m_t^2/4M_W^2$ for
large $m_t$.  In Fig.~\ref{fig:1}, we show where the $1\sigma$ contour of
$R_b$ lies in the $\delta a_R^b$--$\delta a_L^b$ plane.  Due to the
subtraction of (\ref{eq:alsm}), the origin corresponds to the SM with an
unphysical $m_t=0$.  Other values of $m_t$ are indicated on the figure,
showing the $\sim2\sigma$ disagreement for $m_t\approx174$~GeV.

The forward-backward asymmetry, $A_{\rm FB}^b$, is also sensitive to the
left- and right-handed couplings of the $b$ quark.  From $A_{\rm
FB}^{0,b}={3\over4}{\cal A}_e{\cal A}_b$, we find
\begin{equation}
A_{\rm FB}^{0,b}=(A_{\rm FB}^{0,b})^{\rm SM}+ (A_{\rm FB}^{0,b})^{\rm SM}
{\alpha_*\over\pi s_*^2}{a_L^ba_R^b\over (a_L^b)^4-(a_R^b)^4}
\left[a_R^b\delta a_L^b-a_L^b\delta a_R^b\right]\ .
\label{eq:afb}
\end{equation}
Since $a_L^b$ is about five times larger than $a_R^b$, $R_b$ and
$A_{\rm FB}^b$ give complimentary information on the couplings --- $R_b$
is sensitive to $\delta a_L^b$ whereas $A_{\rm FB}^b$ is sensitive to
$\delta a_R^b$.  In addition, $R_b$, unlike $A_{\rm FB}^b$, is almost
insensitive to oblique corrections%
%
%\footnote{The oblique correction to $R_b$ is $\sim
%0.02 {\alpha_*\over 2 \pi s_*^2} \delta s_*^2 \sim 0.0001 \delta s_*^2$
%where $\delta s_*^2$, as a function of $S$ and $T$, is $\sim 1$.}.
%
\footnote{In terms of $S$ and $T$ \cite{peskin}, the oblique correction to
$R_b$ is given by $\delta R_b\approx0.00014S-0.00008T$.}.
Thus at present, we
focus only on $R_b$, although we expect improved measurements of the
forward-backward asymmetry as well as $A_{\rm FB}^{\rm pol}(b)$ from SLC
in the future.

We now address the issue of how large can $\delta a_{L,R}^b$ become.  Since
we have separated out an explicit loop factor of $\alpha_*/4\pi s_*^2$,
both $\delta a_L^b$ and $\delta a_R^b$ are generically of order 1.  From
Eq.~(\ref{eq:rb}), we note that $\delta a_L^b$ will affect $R_b$ at about
the 1\% level --- the same as the present experimental precision.  Thus
the $\delta a_L^b$ contribution to $R_b$ is of roughly the same importance
as that of the oblique parameters $S$, $T$ and $U$ to other electroweak
observables.  $\delta a_R^b$, on the other hand, has a much smaller
effect on $R_b$.  At the same time, in most models, additional right-handed
currents are either suppressed or not present.  Thus $\delta a_R^b$ is often
of much less importance.

While the parameter $\epsilon_b$ \cite{altarelli} has been introduced to
describe the $Z\to b\overline{b}$ vertex, it is defined in relation to the
partial width, $\Gamma(Z\to b\overline{b})$ instead of the ratio $R_b$.
Thus the extraction of $\epsilon_b$ from experimental data is more sensitive
to the determination of $\alpha_s(M_Z)$ as well as the treatment of oblique
corrections.  Nevertheless, assuming $\delta a_R^b\approx0$, $\epsilon_b$ is
related to the vertex correction by $\epsilon_b\approx-{\alpha_*\over2\pi
s_*^2}\delta a_L^b$.

\section{New scalars and $\delta a_{L,R}$ --- the 2HD model}
The 2HD model is one of the simplest extensions of the SM.  In this model,
an additional Higgs doublet is introduced, leading to additional vertex
corrections with physical charged and neutral Higgs loops.
In the 2HD model, separate Higgs doublets give masses to the up- and
down-type quarks.
Due to large couplings to the $t$-quark in the vertex, $R_b$ has been
used to rule out most of the small $\tan\beta$
region of parameter space \cite{hollik,park} where $\tan\beta=v_2/v_1$ with
$v_2$ and $v_1$ giving masses to up- and down-type quarks respectively.

Although we include the 2HD model in our discussion, we also allow for more
general scalar interactions.  In the simplest case, we add a single new
scalar, $\phi$ with arbitrary isospin and electric charge, to the SM.
In order to give a direct contribution to the $Z$-$b$-$\overline{b}$ vertex,
$\phi$ must couple to the $b$-quark.  We first
investigate $\phi$ with a left-handed coupling
\begin{equation}
{\cal L}_Y = g\lambda_L\overline{b}_L\phi F_R+\ldots+{\rm H.c.}\ ,
\end{equation}
where $F_R$ may be either an ordinary or a new right-handed quark.  While
isospin and electric charge must be conserved in the above interaction,
we only need the
explicit term written above (in general, $\phi$ may be a member of a larger
multiplet, filling out some $SU(2)\times U(1)$ representation).  Let
$T_3(\phi)$ and $Q(\phi)$ denote the third component of isospin and charge
of the new scalar and similarly $\{T_3(F_L),T_3(F_R)\}$ and $Q(F)$ for the
quark $F$.  In general we allow $F_L$ and $F_R$ to transform under
different representations of $SU(2)$ so we may accommodate both vector and
chiral particles%
\footnote{By chiral, we mean chiral under the SM gauge group.}.
By isospin conservation, the above quantities must
satisfy the relation $T_3(b)=T_3(\phi)+T_3(F_R)$, as indicated in
Fig.~\ref{fig:2}a.  Note that when $SU(2)_L\times U(1)_Y$ is broken,
scalars carrying different isospin may mix.  Although we avoid such
cases, they may be treated similarly without much further complication.

The $b$ vertex correction due to the scalar $\phi$ may easily be evaluated.
We find
\begin{eqnarray}
\delta a_L^b(\phi)=\lambda_L^2[(T_3(b)-Q(b)s_*^2)\Theta
-(T_3(\phi)&&-Q(\phi)s_*^2)(\Theta +\Psi)\nonumber\\
&&+(T_3(F_R)-T_3(F_L))\Delta](M_Z^2;M_\phi^2,m_F^2)\ ,
\label{eq:scalarL}
\end{eqnarray}
where the functions $\Theta$, $\Psi$ and $\Sigma$ are given in terms of
finite combinations of Passarino-Veltman functions \cite{hvpv} by
\begin{eqnarray}
\Theta(q^2;M^2,m^2)&=&B_1(0;m^2,M^2)
+[2C_{24}-{\textstyle{1\over2}}-m^2C_0+q^2(C_{22}-C_{23})]
(0,q^2,0;M^2,m^2,m^2)\nonumber\\
\Psi(q^2;M^2,m^2)&=&-B_1(0;m^2,M^2)-2C_{24}(0,q^2,0;m^2,M^2,M^2)\nonumber\\
\Delta(q^2;M^2,m^2)&=&m^2C_0(0,q^2,0;M^2,m^2,m^2)\ .
\label{eq:sfunc}
\end{eqnarray}
For degenerate masses present in the $C$ functions, use of Passarino-Veltman
identities allows us to rewrite $\Theta$ and $\Psi$ to give
\begin{eqnarray}
\Theta(q^2;M^2,m^2)
&=&q^2[C_{12}+2C_{22}-C_{23}](0,q^2,0;M^2,m^2,m^2)\nonumber\\
\Psi(q^2;M^2,m^2) &=&q^2[2C_{22}-C_{23}](0,q^2,0;m^2,M^2,M^2)\ ,
\end{eqnarray}
showing that they vanish in the limit $q^2\to0$.  This should come at no
surprise since it is a consequence of the vector Ward identity.  On the
other hand, $\Delta$ is non-vanishing in this limit, and for small $q^2$
has the expansion $\Delta(q^2;M^2,m^2)=\Delta_0(x)+O(q^2/M^2)$ where
$x=m^2/M^2$ and
\begin{equation}
\Delta_0(x)= {x\over1-x}+{x\over(1-x)^2}\log x\ .
\end{equation}

In the same spirit as the $STU$ parameters, we may consider an expansion in
inverse powers of $M_\phi^2$, the new physics scale.  In this case, since
$\Theta$ and $\Psi$ are suppressed relative to $\Delta$ by a factor of
$q^2/M_\phi^2=M_Z^2/M_\phi^2$, the lowest order expression for $\delta
a_L^b(\phi)$ becomes simply
\begin{equation}
\delta a_L^b(\phi)=\lambda_L^2(T_3(F_R)-T_3(F_L))\Delta_0(m_F^2/M_\phi^2)
+\cdots
\label{eq:dals}
\end{equation}
This expression allows us to make a few observations.  First of all,
since it vanishes when $F_L$ and $F_R$ carry the same isospin, chiral
fermions in the loop are necessary in order to get large shifts in $\delta
a_L^b$ (and hence $R_b$).  For vector fermions, the Ward identity ensures
the vertex correction is suppressed by the factor $M_Z^2/M_\phi^2$, giving
another example of the decoupling theorem.  Secondly, $\Delta_0$ vanishes
in the limit $m_F\to 0$, so contributions from light fermions are suppressed
as well.  Both conditions are necessary in order for the new physics to
generate sizable effects in the vertex.  Finally, since $-1<\Delta_0(x)\le0$
for all $x$, the sign of $\delta a_L^b(\phi)$ is completely determined
(to lowest order), and hence the direction of the shift in $R_b$ is fixed
simply by the isospins of the quark $F$.

For a scalar $\chi$ that interacts with right-handed $b$-quarks, similar
expressions may be derived for $\delta a_R^b$.  In this case, as shown in
Fig.~\ref{fig:2}b, isospin conservation demands $T_3(\chi)=-T_3(F_L)$.  The
result is similar to (\ref{eq:scalarL}) and (\ref{eq:dals}), but with
$L\leftrightarrow R$:
\begin{eqnarray}
\delta a_R^b(\chi)&=&\lambda_R^2[(-Q(b)s_*^2)\Theta
-(T_3(\chi)-Q(\chi)s_*^2)(\Theta+\Psi)+(T_3(F_L)-T_3(F_R))\Delta]
(M_Z^2;M_\chi^2,m_F^2)\nonumber\\
&=&\lambda_R^2(T_3(F_L)-T_3(F_R))\Delta_0(m_F^2/M_\chi^2)+\cdots
\label{eq:scalarR}
\end{eqnarray}
Note that $\chi$ may be the same scalar as $\phi$, provided
$T_3(b)=T_3(F_R)-T_3(F_L)$ for consistency among the left and right Yukawa
interactions.  This is indeed the case for the charged Higgs loop in the
2HD model where $F=t$.

Using these results, we now examine the 2HD model in greater detail.
Focusing only on the charged Higgs, $H^+$, the vertex correction $\delta
a_{L,R}^b(H^+)$ may be calculated from (\ref{eq:scalarL}) and
(\ref{eq:scalarR}) using $\lambda_L=m_t\cot\beta/\sqrt{2}M_W$ and
$\lambda_R=m_b\tan\beta/\sqrt{2}M_W$.  For small values of $M_{H^+}$,
the exact expressions need to be used since the expansion factor
$M_Z^2/M_{H^+}^2$ is not sufficiently small.  Nevertheless, the vertex
is typically still dominated by the isospin splitting term
\begin{equation}
\delta a_{L,R}^b(H^+)\approx\mp{\textstyle{1\over2}}\lambda_{L,R}^2
\Delta_0(m_t^2/M_{H^+}^2)\ ,
\end{equation}
where the top sign corresponds to $\delta a_L^b$.
Because of the Yukawa couplings, the left- (right)-handed
interactions dominate for small (large) $\tan\beta$.  This is evident in
Fig.~\ref{fig:1} where we have shown how $\delta a_L^b$ and $\delta a_R^b$
are shifted in the 2HD model relative to the SM with $m_t=150$~GeV.
Note that, regardless of $\tan\beta$, the prediction for $R_b$ (from the
charged Higgs) is always {\it decreased} compared to the SM.

This figure also shows that since $R_b$ is less sensitive to changes in
$\delta a_R^b$ it more easily rules out the small $\tan\beta$ region of the
2HD model.  Furthermore, for large $\tan\beta$, the neutral Higgs
couplings will become important, and the neutral Higgs sector will also play
a role \cite{denner}.
In principle, the $H^+$ loop also generates dipole form factors
proportional to $\lambda_L\lambda_R= m_b m_t/2M_W^2$.  However, as this is
independent of $\tan\beta$, the dipole terms have no enhancement and may
safely be neglected.
In order to demonstrate how strong the constraints are, we show
the 99\%C.L.\ excluded region for several values of $m_t$ in Fig.~\ref{fig:3}.

While the 2HD model essentially describes the Higgs sector of the minimal
supersymmetric standard model (MSSM), the $Z$-$b$-$\overline{b}$ vertex in
the latter model picks up additional contributions from both neutralino and
chargino loops \cite{djouadi,boulware}.  Since these contributions may have
either sign \cite{boulware}, the MSSM, unlike the 2HD model, actually allows
for predictions of $R_b$ in closer agreement with experiment \cite{wells}.

\section{New gauge bosons}
We now turn to the effect of new gauge bosons on the $b$ vertex.  In a
similar vein to the previous section, we consider the addition of a new
gauge boson, $V_L$ or $V_R$,
with either a left- or a right-handed coupling to the $b$,
$V$-$b$-$\overline{F}$.  When $V_L=W$ and $F=t$, this reproduces the SM vertex
correction.  However, we again allow
for the possibility that $F$ is a new quark.  For example, in the
$SU(3)\times U(1)$ model of \cite{pisanoframpton}, the ordinary quark
doublet, $(t,b)$, is extended by the addition of a charge $5/3$ quark, $T$,
to fill out a $SU(3)$ anti-triplet $(b,t,T)$.

Working in 't-Hooft-Feynman gauge, the result for a $V_L$ is given
by
\begin{eqnarray}
\delta a_L^b(V_L)=[(T_3(b)-Q(b)s_*^2){\textstyle{1\over2}}
\Phi+(T_3(V)-Q(V)&&s_*^2)[B_0(0;M_V^2,M_V^2)-{\textstyle{1\over2}}
(\Phi+\Lambda)]\nonumber\\
+&&(T_3(F_R)-T_3(F_L))\Xi](M_Z^2;M_V^2,m_F^2)\ .
\label{eq:vectorL}
\end{eqnarray}
Since isospin is conserved at the vertex, we must ensure
$T_3(b)=T_3(F_L)+T_3(V)$.  Once again, we have ignored isospin mixing that,
however, would generally arise once $SU(2)\times U(1)$ is broken.
The functions are given by
\begin{eqnarray}
{\textstyle{1\over2}}\Phi(q^2;M^2,m^2)&=&
(1+{\textstyle{x\over2}})\Theta(q^2;M^2,m^2)-q^2[C_0+C_{11}]
(0,q^2,0;M^2,m^2,m^2)\nonumber\\
{\textstyle{1\over2}}\Lambda(q^2;M^2,m^2)&=&
(1+{\textstyle{x\over2}})\Psi(q^2;M^2,m^2)+q^2C_{11}
(0,q^2,0;m^2,M^2,M^2)\nonumber\\
&&\qquad\quad-[B_0(q^2;M^2,M^2)-B_0(0;M^2,M^2)]\nonumber\\
\Xi(q^2;M^2,m^2)&=&2m^2C_0(0,q^2,0;m^2,M^2,M^2)
-m^2C_0(0,q^2,0;M^2,m^2,m^2)\nonumber\\
&&\qquad\quad+{\textstyle{x\over2}}[\Theta+\Psi+\Delta](q^2;M^2,m^2)\ .
\label{eq:wfunc}
\end{eqnarray}
As before, $x=m^2/M^2$.  The terms proportional to $x$ come from the
would-be Goldstone boson in 't Hooft-Feynman gauge as can be seen from
comparison with (\ref{eq:scalarL}) and (\ref{eq:sfunc}).

We note that the vertex is finite except for the universal piece arising
from $B_0(0;M_V^2,M_V^2)$ in the non-abelian term.  This is a well known
situation and is removed by a counterterm of a similar form in the on-shell
scheme \cite{lynn,hollik2,passarino,kuroda}.  The treatment is similar
for the * scheme although the subtraction is instead the momentum dependent
term $B_0(q^2;M_V^2,M_V^2)$ \cite{kennedy}.

Taking $F=t$ in (\ref{eq:vectorL}) reproduces the SM top contribution
to the $Z\to b\overline{b}$ vertex \cite{bernabeu,beenakker}.  In
particular, these three vertex functions have been given before in
\cite{lynn}, although in a form reduced to the elementary scalar integrals
$B_0$ and $C_0$.  Prior to numerical evaluation of these functions, we prefer
the above expressions both because of their conciseness and because
they explicitly demonstrate the vanishing of $\Phi$ and $\Lambda$ as
$q^2\to0$.  In the limit $m_F\to0$, we verify that both $\Phi$ and $\Lambda$
reduce to the well known expressions for the vertex in the massless limit
\cite{kennedy,lynn,grzadkowski}.

Similar to the scalar case, this shows that in order for a new gauge
interaction to generate large vertex corrections, the internal fermion,
$F$, must be both massive and chiral.  The vanishing of $\Xi$ as $m_F\to0$
can be understood through helicity conservation in the massless limit since
its coefficient in (\ref{eq:vectorL}) indicates that this term arises from
the difference in isospin between $F_L$ and $F_R$.  Expanding to lowest
order in $q^2/M_V^2$ gives
\begin{equation}
\delta a_L^b(V_L)=(T_3(F_R)-T_3(F_L))\Xi_0(m_F^2/M_V^2)+\cdots\ ,
\end{equation}
where
\begin{eqnarray}
\Xi_0(x)={x(-6+x)\over2(1-x)}-{x(2+3x)\over2(1-x)^2}\log x \ .
\end{eqnarray}
We note that $\Xi_0(x)<0$ for $x\agt0.1$ with $\Xi_0(x)\sim-x/2$ for
large $x$.  Therefore the direction of the shift in $R_b$ is again
determined by the isospins of $F$, at least for the interesting
case when $F$ is heavy.  For the SM, we take $F=t$ and find
$\delta a_L^b(W)\sim x/4$, leading to the asymptotic behavior
$\epsilon_b\sim -G_Fm_t^2/4\sqrt{2}\pi^2$ \cite{altarelli}.

For a $V_R$, we interchange left- and
right-handed couplings and find
\begin{eqnarray}
\delta a_R^b(V_R)=[(-Q(b)s_*^2){\textstyle{1\over2}}
\Phi +(T_3(V)-Q(V)s_*^2)&&[B_0(0;M_V^2,M_V^2)-{\textstyle{1\over2}}
(\Phi +\Lambda)]\nonumber\\
&&+(T_3(F_L)-T_3(F_R))\Xi](M_Z^2;M_V^2,m_F^2)\ ,
\end{eqnarray}
where now the condition $T_3(V_R)=-T_3(Q_R)$ must be satisfied.
As an example of a right-handed interaction, consider the left-right
symmetric (LRS) model \cite{pati,mohapatra,senjanovic}.  There are two
charged gauge bosons in this model, $W_L$ and $W_R$, which may mix with
mixing angle $\zeta$ to form
the mass eigenstates $W_{1,2}$.  In general both $W_1$ and $W_2$ have left-
and right- handed couplings.  However, as the mixing is constrained to be
small \cite{jodidio,langacker}, $W_1$ is mostly $W_L$ and similarly for
$W_2$.  In this case, the contributions from $W_{1,2}$ separate with
$\delta a_L^b(W_1)$ identical to the SM case.  For $W_2$, on the other
hand, although $F=t$ is chiral, it turns out that $\delta a_R^b(W_2)$ is
small since in this case $m_t<M_{W_2}$ \cite{langacker} and hence $\Xi_0$ is
suppressed.

While we have focused on the contributions to $a_{L,R}$, the $W_L$-$W_R$
mixing will induce dipole form factors at the vertex.  This
increases $R_b$ by
\begin{eqnarray}
\frac{\delta R_b}{R_b}&=&(1-R_b)
\frac{\alpha_*}{4\pi s_*^2}|\zeta|^2 \frac{m^2_t/M_W^2}
{(a_L^b)^2+(a_R^b)^2} \nonumber\\
&\sim& 0.05|\zeta|^2\ ,
\end{eqnarray}
which is negligibly small%
\footnote{This mixing effect, however, is important in the decay
$b\rightarrow s \gamma$ \cite{chobabu}.}.
Thus, unlike the 2HD model, $R_b$ does not provide any additional
constraints for the charged gauge boson mixing in the LRS model.

% For example, with $m_t=150$~GeV we find $\delta a_R^b(W_2)=-0.066$ for
% $M_{W_2}=300$~GeV.

\section{Summary}
It is usually assumed that the dominant effects of new physics show up only
obliquely at the current energy scales.  We have examined the validity of
this assumption in more detail for the case of the $b$ vertex.  In
particular, we have considered the effects of both new scalars and new gauge
bosons on the vertex corrections, $\delta a_{L,R}^b$.  By using a
model-independent approach, we find that in general two conditions must be
satisfied by the new particles in order for the vertex correction to compete
with the effects of the oblique corrections.  First of all, the fermion in
the vertex must be chiral, and secondly it must be massive compared to the
boson in the loop.  These conditions are required to avoid the vector Ward
identity which would otherwise constrain the corrections to be small.

Both $R_b$ and $A_{FB}^b$ may be used to constrain the $b$ vertex
corrections.  Both observables are complimentary since the former is mostly
sensitive to $\delta a_L^b$ while the latter is mostly sensitive to $\delta
a_R^b$.  However $R_b$ is especially useful as most of the oblique
effects are cancelled in the ratio, thus allowing us to study the vertex
independently of the oblique corrections that undoubtably arise when new
physics is present.

We have applied the general results to both the 2HD and the LRS model.  For
the 2HD model, we find strong restrictions on the small $\tan\beta$ regime,
but for the LRS model no useful constraints are obtained.  While the top is
present in the loop in both cases, the contrast is mainly due to the behavior
of the Yukawa coupling in the 2HD model, $\lambda_L\sim \cot\beta$, which
enhances the contribution for small $\tan\beta$.

A more careful treatment, including the effects of isospin mixing may be
undertaken in specific models.  We have done this for the $SU(3)\times U(1)$
model \cite{pisanoframpton} and found that the effects are quite small
since the new quark $T$ is a $SU(2)\times U(1)$ vector \cite{liung}.
While isospin
mixing plays a role, the additional terms are proportional to both
$M_W^2/M_Y^2$ and the mass splitting between the new gauge bosons, $Y$, and
the new scalars.  Thus, we feel these general results without the
incorporation of isospin splitting are sufficient to present a good
understanding of new physics and the $b$ vertex.

When both conditions for large vertex corrections are satisfied, the sign of
the contribution to $\delta a_{L,R}^b$ is fixed by the isospin of the fermion
in the loop.  This may provide a clue to what possible new physics may arise
in order to bring $R_b$ into closer agreement with experiment.  For instance,
while the 2HD model always predicts a smaller $R_b$ than the SM, the
additional contributions in the MSSM may lead corrections of either sign
\cite{boulware,wells}.  This is also the case for $b\to s\gamma$ in these
two models \cite{barbieri}.  Of course it remains to be seen whether the
experimental discrepancy in $R_b$ will hold up in the future or not.

%%\section*{Acknowledgements}
\bigskip
This work was supported in part by the National Science Foundation under Grant
No.~PHY-916593, and by the Natural Science and Engineering Research Council
of Canada.

\begin{figure}
\caption{The $1\sigma$ contour for $R_b$ in the $\delta a_L^b$--$\delta
a_R^b$ plane.  The SM predictions with a heavy top are given by the solid
line with $\delta a_R^b=0$.  Also included in the figure are the small and
large $\tan\beta$ behavior of the vertex corrections in the 2HD model in
the case where $m_t=150$~GeV.
}
\label{fig:1}
\end{figure}

\begin{figure}
\caption{Yukawa interactions for new scalars $\phi$ and $\chi$.}
\label{fig:2}
\end{figure}

\begin{figure}
\caption{The 99\% C.L.\ excluded region in the $M_{H^+}$---$\tan\beta$ plane
for the 2HD model.  Note that the SM with $m_t>200$~GeV is already
excluded at 99\%C.L.\ by $R_b$, thus excluding the 2HD model with such a
heavy top as well.}
\label{fig:3}
\end{figure}

%%%%%%%%%%%%%%%%%%%%%
% remove the following comment if you don't want figures included
%%%%%%%%%%%%%%%%%%%%%
%\end{document}
%%%%%%%%%%%%%%%%%%%%%

\newpage
\input psfig
\centerline{\psfig{figure=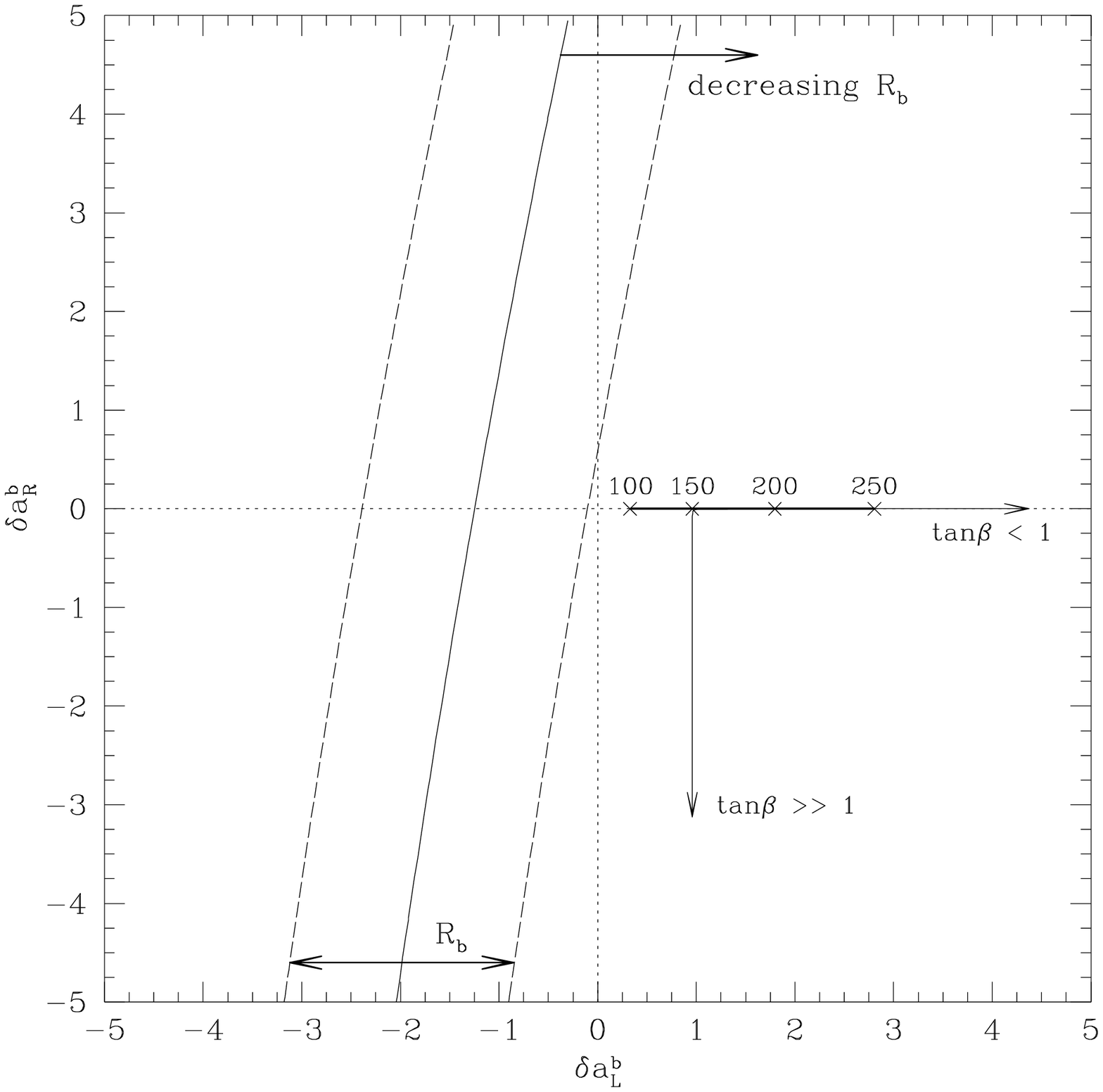}}
\centerline{\bf Figure 1}

\newpage
\input feynman
\bigphotons
\ \vskip3in
\begin{center}
\begin{picture}(45000,20000)
\THICKLINES
%
% Figure 2a
%
\drawline\fermion[\E\REG](1500,7000)[8500]
\drawarrow[\LDIR\ATTIP](\pmidx,\pmidy)
\global\advance\pmidy by 500
\global\advance\pmidx by  -800
\put(\pmidx,\pmidy){\large L}
\global\advance\pmidy by -2000
\global\advance\pmidx by -1000
\put(\pmidx,\pmidy){$T_3(b)$}
\global\advance\pfrontx by -1500
\global\advance\pfronty by -300
\put(\pfrontx,\pfronty){\large $b$}
\global\seglength=1200
\global\gaplength=700
\drawline\scalar[\N\REG](\pbackx,\pbacky)[7]
\drawarrow[\LDIR\ATBASE](\pmidx,\pmidy)
\global\advance\pmidx by 500
\global\advance\pmidy by 2000
\put(\pmidx,\pmidy){$T_3(\phi)$}
\global\advance\pbacky by 800
\global\advance\pbackx by -400
\put(\pbackx,\pbacky){\large $\phi$}
\drawline\fermion[\E\REG](\pfrontx,\pfronty)[8500]
\drawarrow[\LDIR\ATTIP](\pmidx,\pmidy)
\global\advance\pmidy by 500
\global\advance\pmidx by -800
\put(\pmidx,\pmidy){\large R}
\global\advance\pmidy by -2000
\global\advance\pmidx by -300
\put(\pmidx,\pmidy){$T_3(F_R)$}
\global\advance\pbackx by 1000
\global\advance\pbacky by -300
\put(\pbackx,\pbacky){\large $F$}
\global\advance\pfronty by -1500
\global\advance\pfrontx by -800
\put(\pfrontx,\pfronty){\large $g\lambda_L$}
\global\advance\pfronty by -5500
\put(\pfrontx,\pfronty){\large (a)}
%
% Figure 2b
%
\drawline\fermion[\E\REG](25000,7000)[8500]
\drawarrow[\LDIR\ATTIP](\pmidx,\pmidy)
\global\advance\pmidy by 500
\global\advance\pmidx by -800
\put(\pmidx,\pmidy){\large R}
\global\advance\pmidy by -2000
\global\advance\pmidx by -3000
\put(\pmidx,\pmidy){$T_3(b_R)=0$}
\global\advance\pfrontx by -1500
\global\advance\pfronty by -300
\put(\pfrontx,\pfronty){\large $b$}
\global\seglength=1200
\global\gaplength=700
\drawline\scalar[\N\REG](\pbackx,\pbacky)[7]
\drawarrow[\LDIR\ATBASE](\pmidx,\pmidy)
\global\advance\pmidx by 500
\global\advance\pmidy by 2000
\put(\pmidx,\pmidy){$T_3(\chi)$}
\global\advance\pbacky by 800
\global\advance\pbackx by -400
\put(\pbackx,\pbacky){\large $\chi$}
\drawline\fermion[\E\REG](\pfrontx,\pfronty)[8500]
\drawarrow[\LDIR\ATTIP](\pmidx,\pmidy)
\global\advance\pmidy by 500
\global\advance\pmidx by -800
\put(\pmidx,\pmidy){\large L}
\global\advance\pmidy by -2000
\global\advance\pmidx by -300
\put(\pmidx,\pmidy){$T_3(F_L)$}
\global\advance\pbackx by 1000
\global\advance\pbacky by -300
\put(\pbackx,\pbacky){\large $F$}
\global\advance\pfronty by -1500
\global\advance\pfrontx by -800
\put(\pfrontx,\pfronty){\large $g\lambda_R$}
\global\advance\pfronty by -5500
\put(\pfrontx,\pfronty){\large (b)}
\end{picture}
\end{center}
\vskip 1in
\centerline{\bf Figure 2}

\newpage
\centerline{\psfig{figure=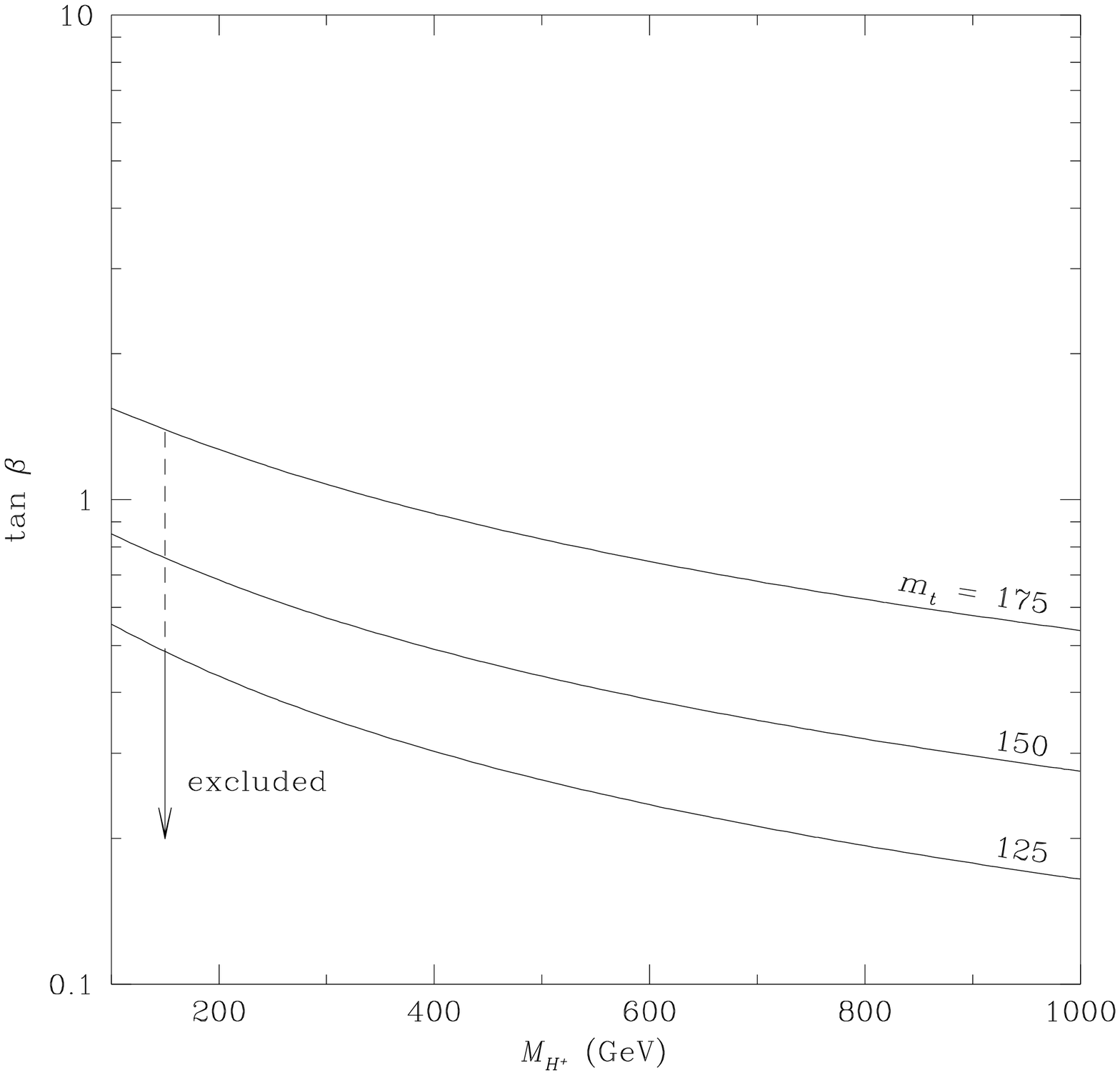}}
\centerline{\bf Figure 3}

\end{document}